\documentclass[pdflatex,iicol,sn-nature]{sn-jnl}

\usepackage{graphicx}%
\usepackage{multirow}%
\usepackage{amsmath,amssymb,amsfonts}%
\usepackage{amsthm}%
\usepackage{mathrsfs}%
\usepackage[title]{appendix}%
\usepackage{xcolor}%
\usepackage{textcomp}%
\usepackage{manyfoot}%
\usepackage{booktabs}%
\usepackage{algorithm}%
\usepackage{algorithmicx}%
\usepackage{algpseudocode}%
\usepackage{listings}%
 \usepackage{indentfirst}

\theoremstyle{thmstyleone}%
\theoremstyle{thmstyletwo}%

\theoremstyle{thmstylethree}%

\begin{document}

\title[Article Title]{Advancing the Control of Low-Altitude Wireless Networks: Architecture, Design Principles, and Future Directions}


\author[1]{\fnm{Haijia} \sur{Jin}}\email{jinhj2024@mail.sustech.edu.cn}

\author*[1]{\fnm{Weijie} \sur{Yuan}}\email{yuanwj@sustech.edu.cn}

\author[1]{\fnm{Jun} \sur{Wu}}\email{wuj2021@mail.sustech.edu.cn}

\author[2]{\fnm{Jiacheng} \sur{Wang}}\email{jiacheng.wang@ntu.edu.sg}
\author[2]{\fnm{Dusit} \sur{Niyato}}\email{dniyato@ntu.edu.sg}

\author[3]{\fnm{Xianbin} \sur{Wang}}\email{xianbin.wang@uwo.ca}

\author[4]{\fnm{George K.} \sur{Karagiannidis}}\email{geokarag@auth.gr}

\author[1]{\fnm{Zhiyun} \sur{Lin}}\email{linzy@sustech.edu.cn}

\author[5]{\fnm{Yi} \sur{Gong}}\email{gongy@sustech.edu.cn}

\author[6]{\fnm{Dong In} \sur{Kim}}\email{dongin@skku.edu}

\author[7]{\fnm{Athina} \sur{Petropulu}}\email{athinap@soe.rutgers.edu}

\author[8]{\fnm{Maria Sabrina} \sur{Greco}}\email{m.greco@iet.unipi.it}

\author[9]{\fnm{Abbas} \sur{Jamalipour}}\email{a.jamalipour@ieee.org}

\author[10]{\fnm{Sumei} \sur{Sun}}\email{sunsm@i2r.a-star.edu.sg}

\affil*[1]{\orgdiv{School of Automation and Intelligent Manufacturing}, \orgname{Southern University of Science and Technology}, \orgaddress{\city{Shenzhen}, \postcode{518055}, \country{China}}}

\affil[2]{\orgdiv{College of Computing and Data Science}, \orgname{Nanyang Technological University}, \orgaddress{\postcode{639798}, \country{Singapore}}}

\affil[3]{\orgdiv{Department of Electrical and Computer Engineering}, \orgname{Western University}, \orgaddress{\city{London}, \country{Canada}, \postcode{N6A 5B9}}}

\affil[4]{\orgname{Aristotle University of Thessaloniki}, \orgaddress{\city{Thessaloniki}, \country{Greece}}}

 \affil[5]{\orgdiv{Department of Electrical and Electronic Engineering}, \orgname{Southern University of Science and Technology}, \orgaddress{\city{Shenzhen}, \postcode{518055}, \country{China}}}

\affil[6]{\orgdiv{Department of Electrical and Computer Engineering}, \orgname{Sungkyunkwan University}, \orgaddress{\city{Suwon}, \postcode{16419}, \country{South Korea}}}

\affil[7]{\orgdiv{Department of Electrical and Computer Engineering}, \orgname{Rutgers, The State University of New Jersey}, \orgaddress{\city{New Brunswick}, \state{NJ}, \country{USA}}}

\affil[8]{\orgdiv{Department of Information Engineering}, \orgname{University of Pisa}, \orgaddress{\postcode{56122}, \city{Pisa}, \country{Italy}}}

\affil[9]{\orgdiv{School of Electrical and Information Engineering}, \orgname{University of Sydney}, \orgaddress{\city{Sydney}, \postcode{NSW 2006}, \country{Australia}}}

\affil[10]{\orgdiv{ Institute for Infocomm Research}, \orgname{Agency for Science,
 Technology and Research}, \country{Singapore}}



\abstract{This article introduces a control-oriented low-altitude wireless network (LAWN) that integrates near-ground communications and remote estimation of the internal system state. This integration supports reliable networked control in dynamic aerial-ground environments. First, we introduce the network's modular architecture and key performance metrics. Then, we discuss core design trade-offs across the control, communication, and estimation layers. A case study illustrates closed-loop coordination under wireless constraints. Finally, we outline future directions for scalable, resilient LAWN deployments in real-time and resource-constrained scenarios.}

\keywords{LAWN, networked control, wireless communication, remote estimation.}



\maketitle

\section{Introduction}\label{sec1}

\begin{figure}
\centering
\includegraphics[width=1\linewidth]{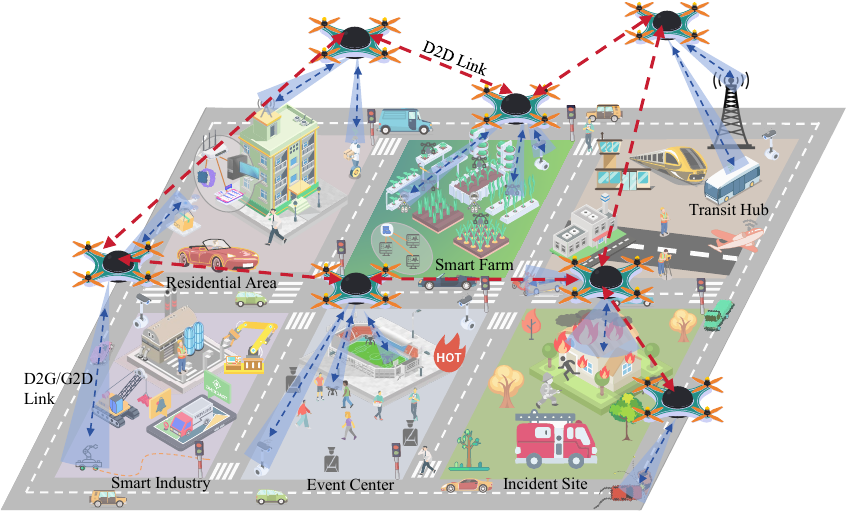}
\caption{Typical scenarios of LAWNs supporting networked control applications. Drones form an adaptive aerial network comprising drone-to-drone (D2D) and drone-to-ground/ground-to-drone (D2G/G2D) links, enabling coordinated communication, control, and remote estimation in diverse low-altitude environments.}
\label{fig:sys}
\end{figure}

The rapid advancement of the Internet of Things (IoT) has expedited the implementation of distributed control systems in industrial, urban, and autonomous contexts. These applications impose stringent requirements on wireless networks, demanding exceptional reliability, ultra-low latency, and adaptability to highly dynamic environments. \cite{IoT_review,8260591}.  In response to these requirements, low-altitude wireless networks (LAWNs) have emerged as a mobility-enhanced communication framework that incorporates drones, near-ground wireless connections, and distributed agents to facilitate delay-sensitive activities \cite{LAWN_IoT,hurst2025uncrewed}.  By leveraging drone mobility and adaptable deployment, LAWNs not only sustain dynamic line-of-sight (LoS) connectivity in intricate surroundings but also enhance wireless coverage in areas with limited infrastructure.  As illustrated in Fig. \ref{fig:sys}, LAWNs facilitate control-critical applications including intelligent transportation, cooperative robotics, and large-scale inspection, where dynamic environments, stringent latency, and evolving agent behaviours require precise control coordination \cite{oubbati2019leveraging,BOURSIANIS2022100187}.

 In LAWN-based systems, the entire control loop, comprising sensing, estimation, decision-making, and actuation, is integrated inside a mobile, entirely wireless framework \cite{zhao2018toward,COISCC1,8766208}. Drones actively monitor ground agents, estimate their states utilising filters, compute commands through techniques such as linear quadratic regulation (LQR) or model predictive control (MPC), and relay control actions via wireless connections \cite{ovsthus2014industrial,zhang2014network,WNCS_review}.  
Thus, LAWN systems empower drones to dynamically prioritise agents with intricate trajectories, restricted observability, or urgent requirements, attaining precise latency adaptation and spatial flexibility that static or pre-coordinated architectures cannot provide \cite{campion2018uav,8385190}.

 Notwithstanding the architectural potential of LAWNs, most of the recent control-oriented LAWNs implementations consider communication, estimation, and control as loosely interconnected modules \cite{11045436}.  The related control techniques frequently presume fixed-delay channels, disregarding latency fluctuations caused by mobility, fading, and obstacles \cite{tsumura2003stabilizability}.  Communication protocols mainly enhance link-level metrics like throughput, energy, or coverage, neglecting the timing guarantees and stability margins crucial for closed-loop control, which leads to delayed or dropped commands that undermine system responsiveness \cite{WNCS_MPC2,11095320}. Similarly,  estimation designs depend on the assumption of complete observations at consistent intervals, which fail in the presence of packet loss and asynchronous updates \cite{huang2020real}.  These deficiencies restrict the capacity of current LAWN systems to function stably and responsively in intricate situations, highlighting the need for an integrated co-design of communication, estimation, and control specifically adapted to LAWN dynamics.

 This article introduces a cohesive framework that incorporates communication and remote estimation into the feedback loop, driven by the necessity for control-oriented integration in LAWNs.  We commence by delineating a modular system architecture and determining various critical performance indicators pertinent to control-driven LAWN operation.  Classical control systems are re-evaluated in distributed, delay-prone contexts to assess their limits in LAWN environments, including susceptibility to estimation inaccuracies, exposure to packet loss, and diminished robustness during asynchronous updates.  This study analyses two fundamental trade-offs to characterise the relationship between wireless restrictions and control behaviour.  We subsequently examine distant state estimation amid packet loss and asynchronous sampling.  An example case study illustrates the simultaneous optimisation of control, estimation, and communication within reliability limitations.  Ultimately, we emphasise prospective research avenues for attaining scalable, adaptive, and resilient networked control utilising LAWNs.

\section{LAWN Architecture}\label{sec2}

\begin{figure*}
\centering
\includegraphics[width=1\linewidth]{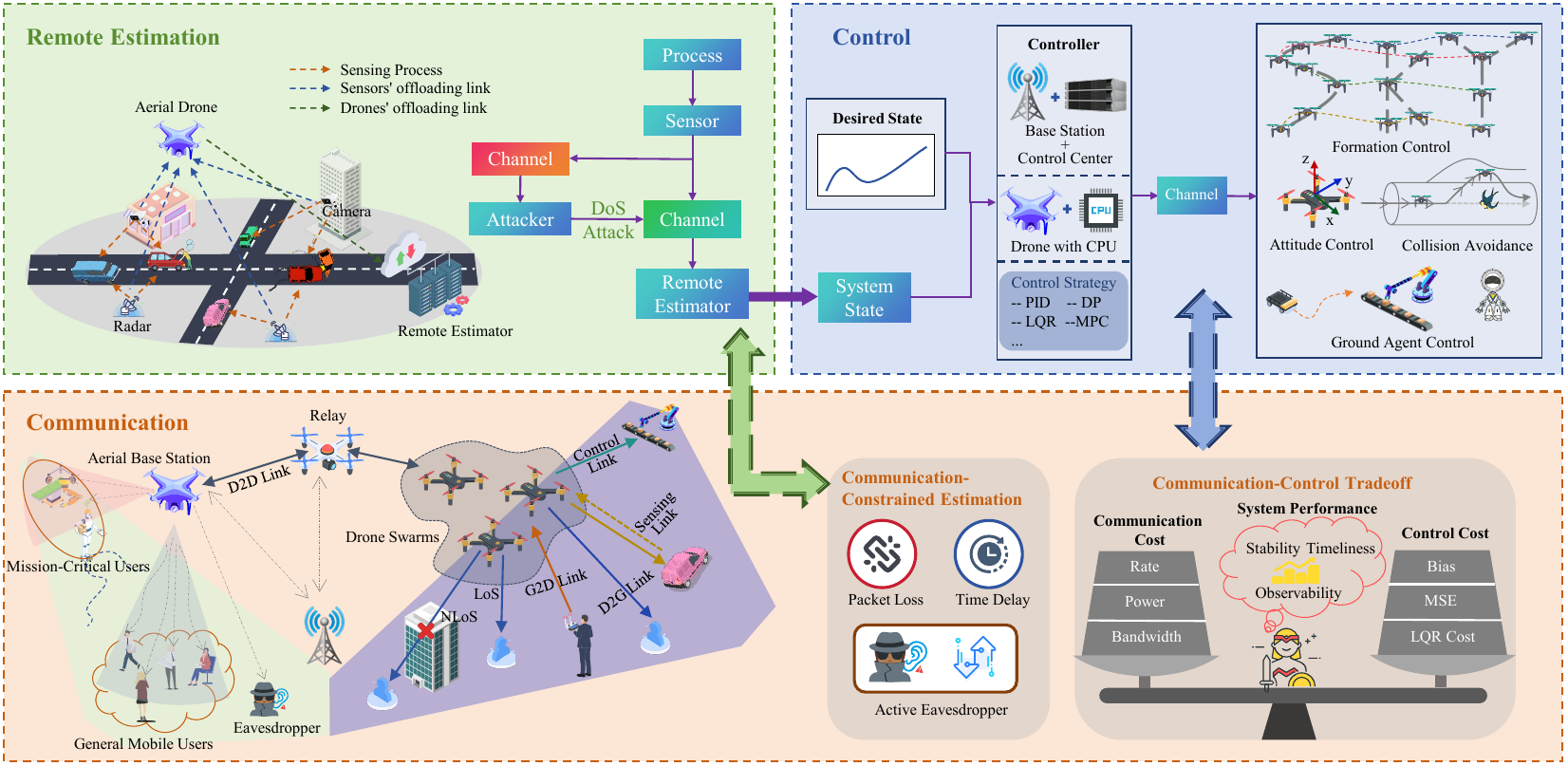}
\caption{A unified architecture of a LAWN system, comprising remote estimation, control, and communication modules.}
\label{fig:function}
\end{figure*}

In this section, we present a conceptual overview of control-oriented LAWNs by outlining their fundamental components and key performance indicators, as depicted in Fig. \ref{fig:function}.

\subsection{Components}

\textbf{Plant:}
The plant represents the dynamic system being controlled, which could be a ground robot, an automated guided vehicle (AGV), or, in certain configurations, a drone. In LAWNs, the plant can consist of multiple distributed agents, such as drones, ground vehicles, or sensor nodes. Each agent exhibits distinct dynamic behaviors and varying levels of communication accessibility. Mathematical models of plant behavior are usually created using stochastic differential equations or discrete-time models.

\textbf{Controller:}
The controller generates control inputs to direct the plant towards a specified state \cite{lewis2012optimal}.  The controller may be located on the drone, a ground node, or a cloud server, contingent upon the system design.  The design must consider both plant dynamics and wireless network limitations, specifically communication delays and link unreliability.

\textbf{Estimator:}
The estimator is responsible for inferring the internal state of the plant based on noisy or delayed measurements \cite{schenato2007foundations}. In LAWN settings, state estimation must account for the stochastic nature of wireless channels, including packet loss, variable delay, and feedback sparsity. Depending on the system architecture, estimators may operate either onboard the drone or at a remote edge controller.

\textbf{Sensor:}
Sensors function as the primary interface between the physical environment and the LAWN architecture \cite{xu2014applications}. They collect raw data essential for state estimation, environmental perception, and control decision-making \cite{wu2025sdrempoweredenvironmentsensingdesign}. Common onboard sensors include cameras, LiDAR, radar, and inertial measurement units (IMUs), each presenting distinct trade-offs in accuracy, sensing range, and latency \cite{10144727}. 

\textbf{Wireless Links:}
Wireless links are a key part of the LAWN communication system, enabling estimation, control, and coordination among drones and ground agents \cite{7984754,9698986,10924402}. In contrast to static networks, these connections demonstrate significant dynamism resulting from channel fading, node movement, and interference.  Signal deterioration results from additive noise (e.g., thermal noise) and multiplicative effects (e.g., fading and Doppler shifts), requiring resilient architectures for estimation and control systems \cite{10791452}.  Three essential connection types are vital to LAWN operation: D2D, D2G, and G2D \cite{yuan2025ground}.  D2D links are influenced by swiftly altering topologies and Doppler shifts, which impede efficient multi-drone coordination \cite{UAV2}.  D2G lines are susceptible to obstruction and elevation-dependent path loss, both of which diminish the reliability of command transmission to ground agents.  G2D lines, tasked with uploading observational data, are affected by channel contention, restricted uplink capacity, and asynchronous sampling periods.

\subsection{Performance Metrics}

A comprehensive assessment of the performance of a LAWN-based control loop system must consider multiple dimensions, including control effectiveness, communication efficiency, and estimation accuracy. This subsection introduces the key performance metrics that characterize the operational quality of control-oriented LAWN systems.

\textbf{Control-Oriented Metrics:}
Control performance is generally measured by the control cost, which reflects the discrepancy between actual and desired system states, along with the effort needed to implement control inputs \cite{jelali2006overview}.  The control cost may also include penalties related to communication, such as energy consumption during transmission.  The stability margin is a crucial metric that indicates the system's capacity to sustain stability amid disturbances and imperfect feedback, especially in scenarios involving communication delays and data losses in wireless environments \cite{kim2017stability}.

\textbf{Communication-Centric Metrics:}
Communication performance in LAWN systems is generally assessed using three primary metrics: latency, reliability, and throughput \cite{chen2010energy}.  Latency refers to the delay experienced in the transmission of communication messages, control commands, or the reception of sensor data.  Reliability denotes the likelihood that transmitted data is accurately received and within specified time constraints \cite{7096295}.  Throughput measures the amount of data effectively transmitted across the network, which is crucial when managing multiple agents or high-resolution sensing data \cite{li2024lowcomplexitydesignirsassistedsecure}.

\textbf{Estimation Quality Metrics:}
The performance of estimation is primarily evaluated through mean squared error (MSE) and update frequency.  The MSE quantifies the average deviation between estimated and actual system states, thereby directly impacting control quality \cite{hahn2004estimation}.  Update frequency denotes the regularity with which the controller acquires new state information \cite{yates2021age}.  High update frequency is essential in dynamic environments, as delays or outdated estimates can severely affect system performance.

 The fundamental elements of LAWNs, including the plant, controller, estimator, sensors, and wireless links, collectively define the structure and functionality of the control loop.  The integration of performance metrics for control, communication, and estimation offers a comprehensive view of the design and operation of control-oriented LAWN systems.

\section{Unified Design in LAWN}\label{sec3}

In LAWNs, control, communication, and estimation are connected due to the underlying wireless architecture, as demonstrated in Fig. \ref{fig:function}.  This section analyses the interaction among these components by investigating representative control strategies, the impact of communication constraints on control performance, and the difficulties of remote state estimation across different system architectures and network conditions.


\begin{table*}[htbp]
\centering
\caption{Performance comparison of control strategies in LAWNs. The terms \textit{Low}, \textit{Moderate}, \textit{High}, \textit{Minimal}, and \textit{Extensive} indicate relative levels of capability or resource consumption, with higher terms reflecting increased effectiveness or computational demand, while \textit{Sub-optimal}, \textit{Near-optimal}, and \textit{Optimal} characterize performance in terms of its proximity to the theoretical optimum.}
\label{table:control_strategies}
\begin{tabular}{|l|c|c|c|c|c|c|}
\hline
\textbf{Strategy}   & \textbf{Efficiency} & \textbf{Robustness} & \textbf{Adaptivity} & \textbf{Complexity} & \textbf{Flexibility} & \textbf{Scalability} \\
\hline
PID   & Sub-optimal & Low & Low & Minimal & Minimal & High \\
\hline
LQR   & Optimal & Moderate & Moderate & Moderate & Low & Moderate \\
\hline
DP    & Optimal & Minimal & Moderate & Extensive & Low & Low \\
\hline
MPC   & Near-optimal & High & High & High & High & Moderate \\
\hline
\end{tabular}
\end{table*}

\subsection{Control Strategies}

Control strategies in LAWN systems must operate under intermittent wireless connectivity, constrained onboard computation, and rapidly evolving environments. Classical techniques, including proportional-integral-derivative (PID), LQR, dynamic programming (DP), and MPC, have been adapted to address these challenges \cite{WNCS_review}. The performance comparison of these control strategies is presented in Table \ref{table:control_strategies}. 

\textbf{PID:}
PID controllers, based on proportional, integral, and derivative feedback, are suitable for low-level tasks such as attitude stabilization and velocity regulation. Their minimal computational cost and scalability make them practical for onboard control of individual drones \cite{knospe2006pid}. In LAWNs, they are often used for stabilizing drone platforms under moderate communication disturbances. However, they lack robustness and adaptivity under severe delays or dynamic link conditions.

\textbf{LQR:}
LQR optimizes a quadratic cost over linear dynamics and is effective for energy-efficient planning and coordinated formation control \cite{cao2009optimal}. In LAWN settings, LQR is commonly applied in structured environments where full-state feedback is available through reliable communication. Its reliance on complete and accurate state information, however, limits its applicability in noisy or partially observable systems.

\textbf{DP:}
DP computes globally optimal policies via Bellman recursion and supports long-term planning under uncertainty \cite{bellman1954theory}. It has been applied in LAWN scenarios for scheduling and global path planning across large areas. Despite its optimality, its low scalability and heavy computation restrict its use in real-time, multi-agent control.

\textbf{MPC:}
MPC performs real-time finite-horizon optimization and is well suited for trajectory tracking and dynamic obstacle avoidance in LAWN-based systems \cite{jin2025predictivecontrollawnjoint}. It offers strong robustness and flexibility under time-varying wireless conditions but requires high computational resources, often necessitating edge or cloud-based deployment.

\subsection{Communication-Control Trade-offs}

Building upon the discussed control strategies, communication quality is a critical factor influencing control performance in LAWN systems. Communication and control are inherently interdependent, and their interplay is shaped by system dynamics and network limitations. Two key trade-offs, the rate-cost trade-off and delay-cost trade-off, illustrate how wireless communication impacts control effectiveness.

\textbf{Rate-Cost Trade-off:} 
The rate-cost trade-off illustrates the inverse relationship between transmission frequency and control performance \cite{10978356}.  In LAWNs, an increase in data rate improves estimation accuracy and control precision; however, this comes with the trade-off of elevated energy consumption and communication overhead \cite{rate_cost_trade_off}.  Balancing communication scheduling with control strategy selection is crucial for optimising performance within resource constraints.

\textbf{Delay-Cost Trade-off:} 
Communication delay has a pronounced effect on control performance, particularly in systems with fast-changing dynamics. Even moderate delays (e.g., 30–50 ms) can significantly impair tracking accuracy and overall control quality \cite{zhang2018rate,FBL11}. In LAWNs, unpredictable transmission delays, arising from link quality and topology changes, necessitate delay-aware strategies such as dynamic scheduling and prioritization of critical data to maintain system stability and responsiveness.


\subsection{Remote Estimation}
In control-oriented LAWNs, remote estimation plays a vital role in enabling informed decision-making among physically and functionally decoupled agents.

\textbf{System Configurations:}
Remote estimation scenarios frequently arise in LAWN environments where the estimator does not have direct access to the target system.  Sensing and estimation functionalities are distributed among distinct agents in these scenarios.  The sensor node, situated near the target, collects measurements and transmits them to the estimator through wireless links \cite{tang2022radix}.  

\textbf{Remote Estimation Mechanisms:}
Various methods are employed for remote estimation in LAWN systems.  Kalman filtering is typically utilised in linear systems characterised by Gaussian noise, whereas adaptations like the extended Kalman filter (EKF) and the unscented Kalman filter (UKF) address moderate nonlinearity \cite{konatowski2016comparison}.  Particle filters (PF) are employed for systems characterised by high nonlinearity or non-Gaussian properties, although they entail significant computational costs.  Distributed estimation, learning-based estimators, and graph-based fusion are utilised in scenarios involving multiple agents, enhancing performance while adhering to bandwidth and latency limitations \cite{pazho2023survey}.

\textbf{Communication-Constrained Estimation:}
Wireless communication limitations significantly affect remote estimation in LAWNs. Packet loss and delays, caused by fading, interference, or mobility, disrupt data flow and hinder timely state updates, leading to estimation errors \cite{sun2017remote}. Additionally, jamming attacks can disrupt critical data transmission, which is countered by secure strategies like randomized scheduling, multi-path redundancy, and secure coding \cite{guo2016optimal}. Adaptive approaches, such as game-theoretic models, help agents adjust communication strategies based on interference, maintaining estimation accuracy in adversarial conditions.

The integration of control, communication, and estimation in LAWN systems necessitates a unified design paradigm. By analyzing the control strategies, characterizing control-communication performance trade-offs, and investigating the challenges of remote estimation under network constraints, this section articulates the need for principled co-design approaches that ensure robust, efficient, and scalable operation in dynamic and distributed environments.

\section{Case Study: Outage-Aware Predictive Control}\label{sec4}

\begin{figure*}
\centering
\includegraphics[width=1\linewidth]{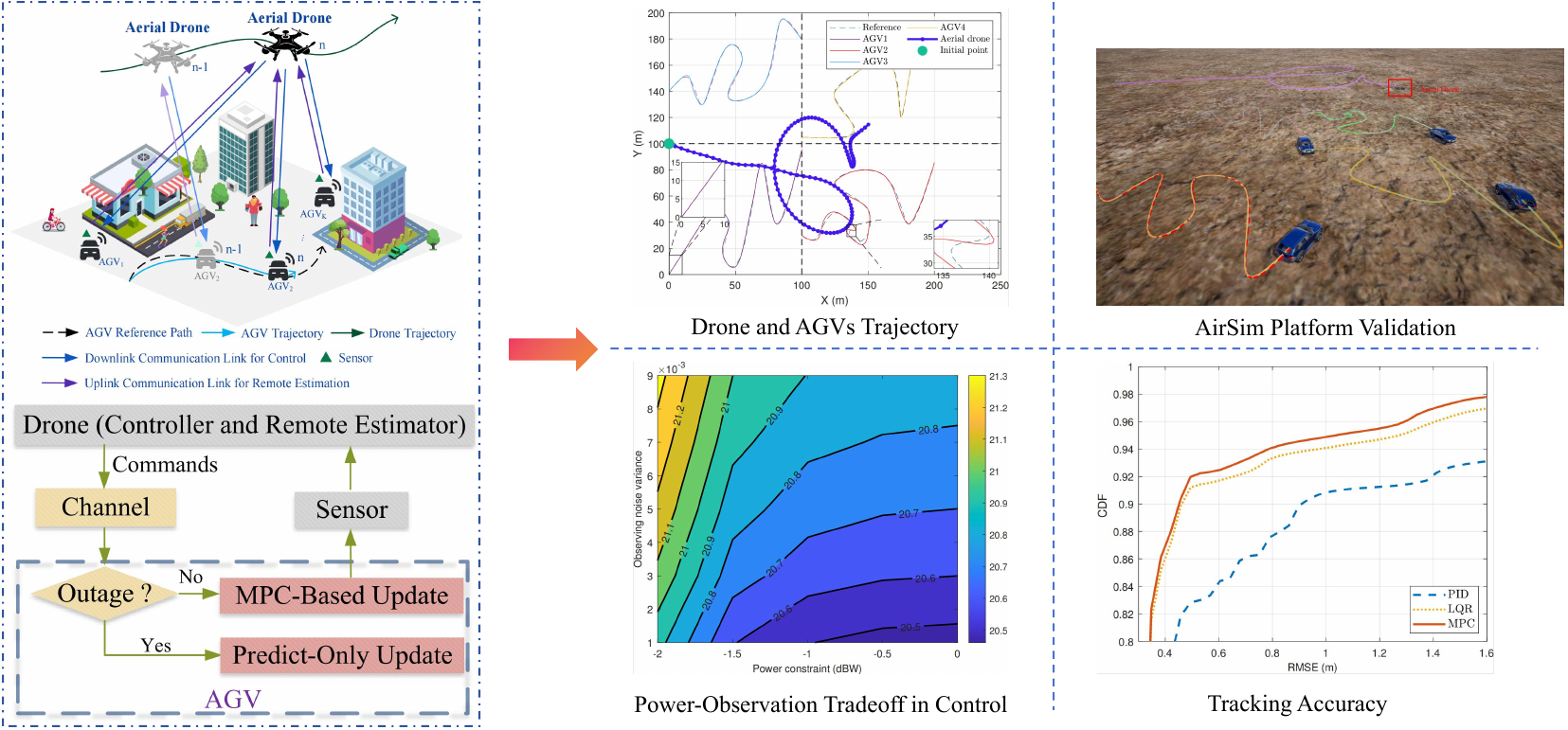}
\caption{LAWN-enabled wireless control case study.
One aerial drone controls multiple AGVs to perform trajectory tracking tasks, integrating remote state estimation and wireless communication.}
\label{fig:sys_mod}
\end{figure*}

This section introduces a representative case study that illustrates the interplay among control, communication, and remote estimation in a multi-agent LAWN system.

\subsection{System Overview} Fig. \ref{fig:sys_mod} depicts a LAWN-enabled wireless control architecture wherein an aerial drone orchestrates multiple AGVs through integrated estimation and control across a time-varying communication network.  Each AGV is equipped with onboard or proximate sensors that relay motion-related data to the drone through uplink channels.  The drone operates as a remote estimator and controller, reconstructing AGV states from incoming data and calculating control actions using an MPC framework.  The actions are subsequently transmitted to the AGVs via downlink channels \cite{jin2025predictivecontrollawnjoint}.  

\subsection{Predictive Control over Wireless Links} The control performance is affected by the design of the control algorithm, the quality of remote state estimation, and the reliability of wireless transmissions.  Inaccurate state reconstruction may result from delayed, noisy, or incomplete observation data, subsequently reducing the efficacy of control decisions.  Moreover, variations in downlink quality may hinder the prompt transmission of control commands to the AGVs.  We implement an adaptive control strategy that utilises the MPC framework to assess future trajectories, integrating the confidence level of current state estimates.  With an increase in estimation uncertainty, the controller implements more conservative measures to maintain system stability \cite{huang2020real}.  AGVs implement a fallback mechanism that utilises locally generated predictions based on previous control inputs, allowing for uninterrupted operation when updated commands are unavailable due to communication or estimation failures \cite{ishii2021mpc}.

\subsection{Results}

\textbf{Simulations:}
We consider a drone-assisted wireless control scenario in a multi-agent LAWN setup, where the drone hovers at a fixed altitude of $50$ m and adjusts its horizontal position to assist four AGVs in motion tracking. The wireless channel is modeled as a probabilistic LoS channel, which accounts for both LoS and non-line-of-sight (NLoS) links \cite{probabilistic_los2}. Moreover, due to the relatively low velocity of the drone, Doppler effects are neglected, as their impact on the communication performance is minimal \cite{xiao2016enabling}. The drone dynamically adjusts its trajectory to assist AGVs during complex maneuvers, maintaining robust tracking performance across varying transmit power levels and sensing conditions, as shown in Fig. \ref{fig:sys_mod}.

\textbf{AirSim Platform-Based Experiment:}
To further evaluate system performance under realistic conditions, the full control-estimation-communication loop is implemented in the AirSim platform \cite{madaan2020airsim}. The simulation environment includes multiple weather effects, with 20\% rain and snow, 30\% road wetness, and 15\% fog. Wind is configured at 2 m/s, 2.5 m/s, and 0 m/s along the X, Y, and Z axes, respectively. As illustrated in Fig. \ref{fig:sys_mod}, the proposed approach maintains stable performance under challenging conditions, demonstrating robustness to physical disturbances and tracking delays.

\section{Future Directions}\label{sec5}

In this section, we outline four key research directions that can enhance the scalability, responsiveness, and robustness of control-oriented LAWNs in dynamic and resource-constrained environments.

\textbf{Security, Resilience, and Fault Tolerance:}
As aerial-ground interaction intensifies, LAWNs face growing risks of eavesdropping, spoofing, and jamming, especially in adversarial environments \cite{jun2025aerial}. The dynamic topology and limited onboard resources hinder the use of conventional security schemes \cite{UAV_IoT2}. Future efforts should explore lightweight encryption, spread-spectrum techniques, and cross-layer authentication tailored to real-time control. In parallel, resilience and fault tolerance are vital to sustain operation under link failures or node compromise \cite{wang2020optimal}. Mechanisms such as adaptive reconfiguration and redundant encoding help preserve system integrity and control performance under disruptions.

\textbf{Delay-Sensitive Control:}
Wireless delays induced by drone mobility, interference, and fluctuating traffic loads can degrade control stability and responsiveness. To satisfy real-time constraints, delay-sensitive design must encompass both the physical and network layers \cite{11045436}. Techniques such as short control frames, Doppler-resilient waveforms, adaptive retransmissions, and latency-aware scheduling can mitigate timing disruptions. A unified treatment of latency across communication and control is essential to ensuring reliable and timely feedback loops.

\textbf{ISAC-Enabled LAWN:}
Integrated sensing and communication (ISAC) has emerged as a promising paradigm for joint hardware reuse, spectrum efficiency, and real-time environmental awareness \cite{liu2022integrated}. In LAWN systems, ISAC can significantly enhance control performance by enabling accurate situational awareness under stringent payload and energy constraints.  Future research should investigate ISAC-based channel estimation, target tracking, and control-aware waveform design to support low-latency, closed-loop feedback in mobility-driven applications.

\textbf{Energy-Aware Design:}
Stable control in LAWNs requires not only timely communication, sensing, and estimation, but also sustained energy availability \cite{7725957}. As drones execute continuous estimation and actuation, energy depletion can lead to degraded responsiveness or control interruptions. Integrating wireless power transfer (WPT) into LAWN operations offers a promising path to support persistent control under mobility constraints \cite{9675011}. Future designs should consider energy availability as a constraint in control policies, ensuring robustness even under limited onboard power.

\textbf{Distributed Coordination:}
In large-scale deployments, centralized control becomes increasingly impractical due to bandwidth limitations and computational bottlenecks.  Control should be decentralised among agents functioning under conditions of partial observability and asynchronous communication \cite{6750042}.  Research should prioritise decentralised estimation, consensus-based coordination, and event-triggered control updates.  The main goal is to attain low-overhead, high-reliability control among spatially distributed drones and ground nodes.

\textbf{AI-Driven Adaptation:}
LAWN systems must be capable of adapting to rapidly changing environments and uncertain dynamics. Although learning-based approaches show great potential, their computational and latency requirements must align with the resource constraints of drone platforms \cite{Wang_Dusit}. Future research should focus on efficient learning models, such as compressed neural networks and federated reinforcement learning, to enable onboard adaptation. Endowing aerial agents with robust, real-time AI capabilities will be essential to sustaining autonomous control in complex and dynamic environments.

\section{Conclusion}\label{sec6}

This article presented a unified perspective on LAWNs, which integrate control, communication, and remote estimation in order to support real-time operations within dynamic environments. It discussed the core architecture, system components, and fundamental trade-offs, and illustrated their interplay through a representative case study. LAWNs offer a scalable infrastructure for networked control systems by enabling responsive and reliable coordination among distributed agents. Looking ahead, advancing the capabilities of control-oriented LAWNs will depend on progress in latency-aware communication, decentralized decision-making, and lightweight adaptive intelligence.

\bibliography{sn-bibliography}

\end{document}